\documentclass[aps,prl,twocolumn]{revtex4}
\usepackage{amssymb,amsbsy,graphicx,times,subfigure,marvosym,color,bm,multirow}
\usepackage[latin1]{inputenc}
\begin{document}

%%%%%%%%%%%%%%%%%%%%%%%%%%%%%%%%%%%%%%%%%%%%%%%%

\title{Topological R\'{e}nyi Entropy after a Quantum Quench}
\author{G\'{a}bor B. Hal\'{a}sz$^{1,2}$}
\author{Alioscia Hamma$^{3,1}$}
\address{$^1$Perimeter Institute for Theoretical Physics, 31 Caroline Street North, Waterloo, Ontario, Canada N2L 2Y5 \\
$^2$Theoretical Physics, Oxford University, 1 Keble Road, Oxford OX1 3NP, United Kingdom \\
$^3$Center for Quantum Information, Institute for Interdisciplinary
Information Sciences, Tsinghua University, Beijing 100084, People's
Republic of China}

%%%%%%%%%%%%%%%%%%%%%%%%%%%%%%%%%%%%%%%%%%%%%%%%

\begin{abstract}

We present an analytical study on the resilience of topological
order after a quantum quench. The system is initially prepared in
the ground state of the toric-code model, and then quenched by
switching on an external magnetic field. During the subsequent time
evolution, the variation in topological order is detected via the
topological R\'{e}nyi entropy of order $2$. We consider two
different quenches: the first one has an exact solution, while the
second one requires perturbation theory. In both cases, we find that
the long-term time average of the topological R\'{e}nyi entropy in
the thermodynamic limit is the same as its initial value. Based on
our results, we argue that topological order is resilient against a
wide range of quenches.

\end{abstract}

%%%%%%%%%%%%%%%%%%%%%%%%%%%%%%%%%%%%%%%%%%%%%%%%

\maketitle

%%%%%%%%%%%%%%%%%%%%%%%%%%%%%%%%%%%%%%%%%%%%%%%%

\emph{Introduction.---}Topologically ordered phases in quantum
many-body systems are of extreme importance in condensed matter
physics \cite{Wen-1} and in quantum information \cite{Kitaev, QC}.
They are novel phases of matter that defy the Landau paradigm of
spontaneous symmetry breaking and possess a robust ground-state
degeneracy that makes them good candidates for quantum memories.
Furthermore, they feature anyonic excitations whose interactions are
of a topological nature and that are therefore less subject to
decoherence \cite{Kitaev, QC}.

Formally, a gapped phase is topologically ordered if and only if it
has a topology-dependent ground-state degeneracy such that $\langle
\Phi | \hat{O} | \Phi' \rangle$ is exponentially small in the system
size for any local operator $\hat{O}$ and any two orthogonal ground
states $| \Phi \rangle$ and $| \Phi' \rangle$. On the other hand, it
has been shown that topological order can be detected in the global
properties of each ground-state wavefunction, without reference to
any other states or the Hamiltonian \cite{Hamma-1, TE, Flammia,
Kim}. More precisely, topological order is revealed by an
entanglement pattern called topological entropy: a universal
correction to the boundary law for the entanglement entropy. Since
this correction is robust against perturbations, it serves as an
effective non-local order parameter for topologically ordered phases
\cite{Hamma-2, Isakov}.

Unfortunately, the entanglement entropy is extremely hard to compute
and its measurement requires complete state tomography \cite{Amico}.
On the other hand, it has been argued that the R\'{e}nyi entropy of
order $2$ also contains substantial information about the universal
properties of a quantum many-body system \cite{Zanardi}. In
particular, the topological pattern of entanglement appears in the
R\'{e}nyi entropy as well \cite{Flammia, Halasz}. Moreover, this
quantity is significantly easier to compute and can in principle be
measured directly \cite{Measurement}.

It is crucial to understand how topological order behaves away from
equilibrium. Since topological order is a property of the
wavefunction only, its presence or absence is well defined for an
arbitrary quantum state, and it can be present in a non-equilibrium
state even if it is absent from the ground state of the system
Hamiltonian. The non-equilibrium properties of quantum many-body
systems are in general extremely fruitful topics in both theoretical
\cite{Polkovnikov} and experimental \cite{Greiner} condensed matter
physics, and they can be studied conveniently in the setting of the
quantum quench: a sudden change in the system Hamiltonian
\cite{Quench}. The quantum quench of a topologically ordered system
was numerically studied in Ref. \cite{Tsomokos}, where they found
that topological order is resilient against certain types of
quenches. The main disadvantage of their method is that it is only
applicable to small system sizes.

In this Letter, we analytically study the behavior of a
topologically ordered system after a quantum quench via the time
evolution of the topological R\'{e}nyi entropy of order $2$. In
particular, we prepare the system in the ground state of the
toric-code model (TCM) \cite{Kitaev} and quench it by switching on
an external magnetic field. By establishing an exact treatment and a
perturbation theory for two different versions of the quench, we
ensure that our results are not confined to the small system sizes
accessible by exact diagonalization \cite{Hamma-2, Tsomokos}.

\begin{figure}[b!]
\centering
\includegraphics[width=5.0cm]{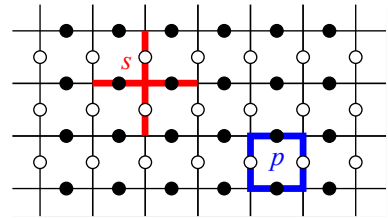}
\caption{(Color online) Illustration of the square lattice with
physical spins on the horizontal (black circles) and the vertical
(white circles) edges. Examples of a star ($s$) and a plaquette
($p$) are included. \label{fig-1}}
\end{figure}

\emph{General formalism.---}We consider the TCM with an external
magnetic field in the $+z$ direction. In this model, there are
$2N^2$ spins on the edges of an $N \times N$ square lattice with
periodic boundary conditions \cite{Kitaev}. The spins on the
horizontal ($h$) and the vertical ($v$) edges experience different
magnetic fields $\lambda$ and $\kappa \lambda$, therefore the
Hamiltonian takes the form \cite{Halasz}
\begin{equation}
\hat{H} (\lambda) = - \sum_s \hat{A}_s - \sum_p \hat{B}_p - \lambda
\sum_{i \in h} \hat{\sigma}_i^z - \kappa \lambda \sum_{i \in v}
\hat{\sigma}_i^z, \label{eq-H-1}
\end{equation}
where the star operators $\hat{A}_s \equiv \prod_{i \in s}
\hat{\sigma}_i^x$ and the plaquette operators $\hat{B}_p \equiv
\prod_{i \in p} \hat{\sigma}_i^z$ belong to stars ($s$) and
plaquettes ($p$) on the lattice containing four spins each (see Fig.
\ref{fig-1}).

In the case of $\lambda = 0$, the system is exactly solvable because
$[\hat{A}_s, \hat{B}_p] = [\hat{A}_s, \hat{H} (0)] = [\hat{B}_p,
\hat{H} (0)] = 0$. The sectors with different sets of expectation
values $\{ A_s \equiv \langle \hat{A}_s \rangle = \pm 1 \}$ and $\{
B_p \equiv \langle \hat{B}_p \rangle = \pm 1 \}$ can be treated
independently. In the ground-state sector with $A_s = +1$ ($\forall
s$) and $B_p = +1$ ($\forall p$), there are four linearly
independent degenerate ground states that are distinguished by the
topological quantum numbers $Z_{1,2} = \pm 1$. These quantum numbers
are expectation values for products of $\hat{\sigma}_i^z$ operators
along horizontal and vertical strings going around the lattice. The
ground state with $Z_1 = Z_2 = +1$ becomes $| 0 \rangle =
\mathcal{N} \prod_{s} (1 + \hat{A}_s) | \Uparrow \, \rangle$, where
$\mathcal{N}$ is a normalization constant and $| \Uparrow \,
\rangle$ is the completely polarized state with $\sigma_i^z \equiv
\langle \hat{\sigma}_i^z \rangle = +1$ ($\forall i$).

In the case of $\lambda > 0$, the system is not exactly solvable in
general because $[\hat{A}_s, \hat{\sigma}_i^z] \neq 0$. On the other
hand, it is true that $[\hat{B}_p, \hat{Z}_{1,2}] = [\hat{B}_p,
\hat{H} (\lambda)] = [\hat{Z}_{1,2}, \hat{H} (\lambda)] = 0$,
therefore the sectors with different values of $B_p$ and $Z_{1,2}$
can be treated independently. In the following, we only consider the
lowest-energy sector with $B_p = +1$ ($\forall p$) and $Z_1 = Z_2 =
+1$. The states within this sector are distinguished by the
expectation values $\{ A_s = \pm 1 \}$, and we introduce a
corresponding representation with quasi-spins $A_s$ located at the
stars \cite{Dusuel}. In this quasi-spin representation, the
quasi-spin $A_s$ is measured by the operator $\hat{A}_s^z \equiv
\hat{A}_s$ and switched by the operator $\hat{A}_s^x$, therefore the
quasi-spin operators $\hat{A}_s^z$ and $\hat{A}_s^x$ satisfy the
standard spin commutation relations. Note that $\hat{\sigma}_i^z =
\hat{A}_s^x \hat{A}_{s'}^x$ for an edge $i$ between two neighboring
stars $s$ and $s'$. Up to an additive constant, the Hamiltonian in
Eq. (\ref{eq-H-1}) becomes
\begin{equation}
\hat{H} (\lambda) = - \sum_s \hat{A}_s^z - \lambda \sum_{\langle
s,s' \rangle \in h} \hat{A}_s^x \hat{A}_{s'}^x - \kappa \lambda
\sum_{\langle s,s' \rangle \in v} \hat{A}_s^x \hat{A}_{s'}^x,
\label{eq-H-2}
\end{equation}
where $\langle s,s' \rangle$ means that the summation is over edges
between neighboring stars $s$ and $s'$. The TCM with external
magnetic field is therefore equivalent to a 2D transverse field
Ising model (TFIM) in which the coupling strengths on the horizontal
and the vertical edges are not equal in general.

When studying the quantum quench, we are interested in the time
evolution of the quantum state $| \Psi(t) \rangle$ after a sudden
change in the Hamiltonian. At time $t = 0$, the system is set up in
the ground state of the initial Hamiltonian $\hat{H} (0)$ such that
$| \Psi(0) \rangle = | 0 \rangle$. At time $t > 0$, the system is
evolved with the quench Hamiltonian $\hat{H} (\lambda)$ and the
state takes the general form $| \Psi(t) \rangle = \exp[-i t \hat{H}
(\lambda)] | 0 \rangle$. To extract any valuable information from
this expression, we need to obtain a full solution of the
Hamiltonian $\hat{H} (\lambda)$. In the following, we consider two
important limits. When $\kappa = 1$, the horizontal and the vertical
coupling strengths are equal, and the equivalent TFIM is the
standard 2D TFIM. When $\kappa \rightarrow 0+$, the vertical
coupling strength vanishes, and the equivalent TFIM factorizes into
$N$ independent 1D TFIM copies along the horizontal chains of the
lattice \cite{Yu}. The $\kappa = 0$ case has an exact solution
available for all values of $\lambda$, while the $\kappa = 1$ case
requires perturbation theory around the exactly solvable point at
$\lambda = 0$.

\emph{Topological R\'{e}nyi entropy.---}The R\'{e}nyi entropy of
order $\alpha$ is a generalization of the usual (von Neumann)
entanglement entropy that characterizes the quantum entanglement
between two complementary subsystems $A$ and $B \equiv
\overline{A}$. It is defined by $S_{\alpha}^{AB} \equiv \log_2
\mathrm{Tr} [\hat{\rho}_A^{\alpha}] / (1 - \alpha)$, where
$\hat{\rho}_A$ is the reduced density operator for $A$. Note that
the usual entanglement entropy is recovered in the special case of
$\alpha = 1$.

\begin{figure}[h!]
\centering
\includegraphics[width=7.5cm]{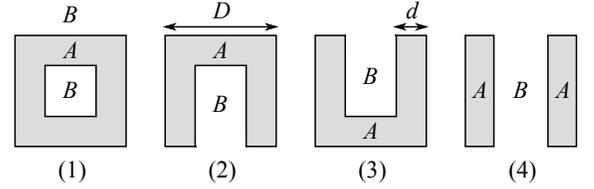}
\caption{Illustration of the subsystems $A$ and $B$ in the four
cases $(m)$ that are used to calculate the topological R\'{e}nyi
entropy. \label{fig-2}}
\end{figure}

The topological R\'{e}nyi entropy is extracted from independent
R\'{e}nyi entropies $S_{\alpha}^{(m)}$ calculated in the four cases
$(m)$ of Fig. \ref{fig-2} as $S_{\alpha}^T \equiv - S_{\alpha}^{(1)}
+ S_{\alpha}^{(2)} + S_{\alpha}^{(3)} - S_{\alpha}^{(4)}$. When
defining $S_{\alpha}^T$ in the standard way \cite{TE}, we take
$S_{\alpha}^{(m)} = S_{\alpha}^{AB}$ between the two thick
subsystems $A$ and $B$. However, it was argued in Ref. \cite{Halasz}
that the presence of the $\mathbb{Z}_2$ lattice gauge structure due
to the constraint $B_p = +1$ ($\forall p$) makes it possible to
substitute the subsystem $A$ with its boundary $\partial A$. In the
following, we use the corresponding modified definition for
$S_{\alpha}^T$ in which we take $S_{\alpha}^{(m)} =
S_{\alpha}^{\partial A, \overline{\partial A}} \equiv
S_{\alpha}^{\partial A}$ between the thin boundary subsystem
$\partial A$ and the rest of the system. The topological R\'{e}nyi
entropy is non-zero if and only if the given state has topological
order. For example, its value is $S_{\alpha}^T = 2$ for the TCM
ground state $| 0 \rangle$ and $S_{\alpha}^T = 0$ for the completely
polarized state $| \Uparrow \, \rangle$. Note that the dimensions
$D$ and $d$ of the subsystems need to be macroscopic such that $D >
d \gg 1$.

The modified definition for the topological R\'{e}nyi entropy is an
immense simplification to our calculations because the reduced
density matrix $\rho_{\partial A}$ is diagonal in the basis of the
physical spins $\sigma_i^z$. It was shown in Ref. \cite{Halasz} that
if the boundary subsystem $\partial A$ consists of $n$ closed loops
with a combined length $L$, the R\'{e}nyi entropy of order $2$ for a
generic state $| \Psi \rangle$ satisfying the gauge constraint $B_p
= +1$ ($\forall p$) is
\begin{eqnarray}
S_2^{\partial A} &=& (L - n) - \log_2 \bigg{[} 1 + \sum_{s_1, s_2}
\langle \Psi | \hat{A}_{s_1}^x \hat{A}_{s_2}^x | \Psi \rangle^2
\nonumber \\
&+& \sum_{s_1, s_2, s_3, s_4} \langle \Psi | \hat{A}_{s_1}^x
\hat{A}_{s_2}^x \hat{A}_{s_3}^x \hat{A}_{s_4}^x | \Psi \rangle^2\ +
\ldots \bigg{]}, \label{eq-S-0}
\end{eqnarray}
where the sum inside the logarithm contains all $2^{L-n}$ possible
products with an even number of quasi-spin operators $\hat{A}_s^x$
chosen from each closed loop of the subsystem $\partial A$.

\emph{Exact treatment.---}In the case of $\kappa = 0$, there are no
interactions between the horizontal chains of the lattice, and the
2D system factorizes into $N$ independent 1D systems in terms of the
quasi-spins $A_s$. Indeed, the 2D Hamiltonian in Eq. (\ref{eq-H-2})
becomes the direct sum of $N$ identical 1D Hamiltonians. If we
consider any horizontal chain and label its $N$ stars with $1 \leq l
\leq N$, the corresponding 1D Hamiltonian is
\begin{equation}
\hat{h} (\lambda) = - \sum_{l=1}^{N} \left( \hat{A}_l^z + \lambda
\hat{A}_l^x \hat{A}_{l-1}^x \right), \label{eq-H-3}
\end{equation}
where the periodic boundary conditions are taken into account by
$A_0 \equiv A_N$. Since the 1D TFIM Hamiltonian in Eq.
(\ref{eq-H-3}) is exactly solvable by means of a standard procedure
\cite{Barouch}, we can determine the exact time evolution of the
R\'{e}nyi entropy after the quantum quench \cite{Supplement}. Note
that despite the factorization into independent 1D chains in terms
of the quasi-spins, this calculation gives the R\'{e}nyi entropy of
a strongly entangled 2D state in terms of the physical spins.

\begin{figure}[t!]
\centering
\includegraphics[width=7.8cm]{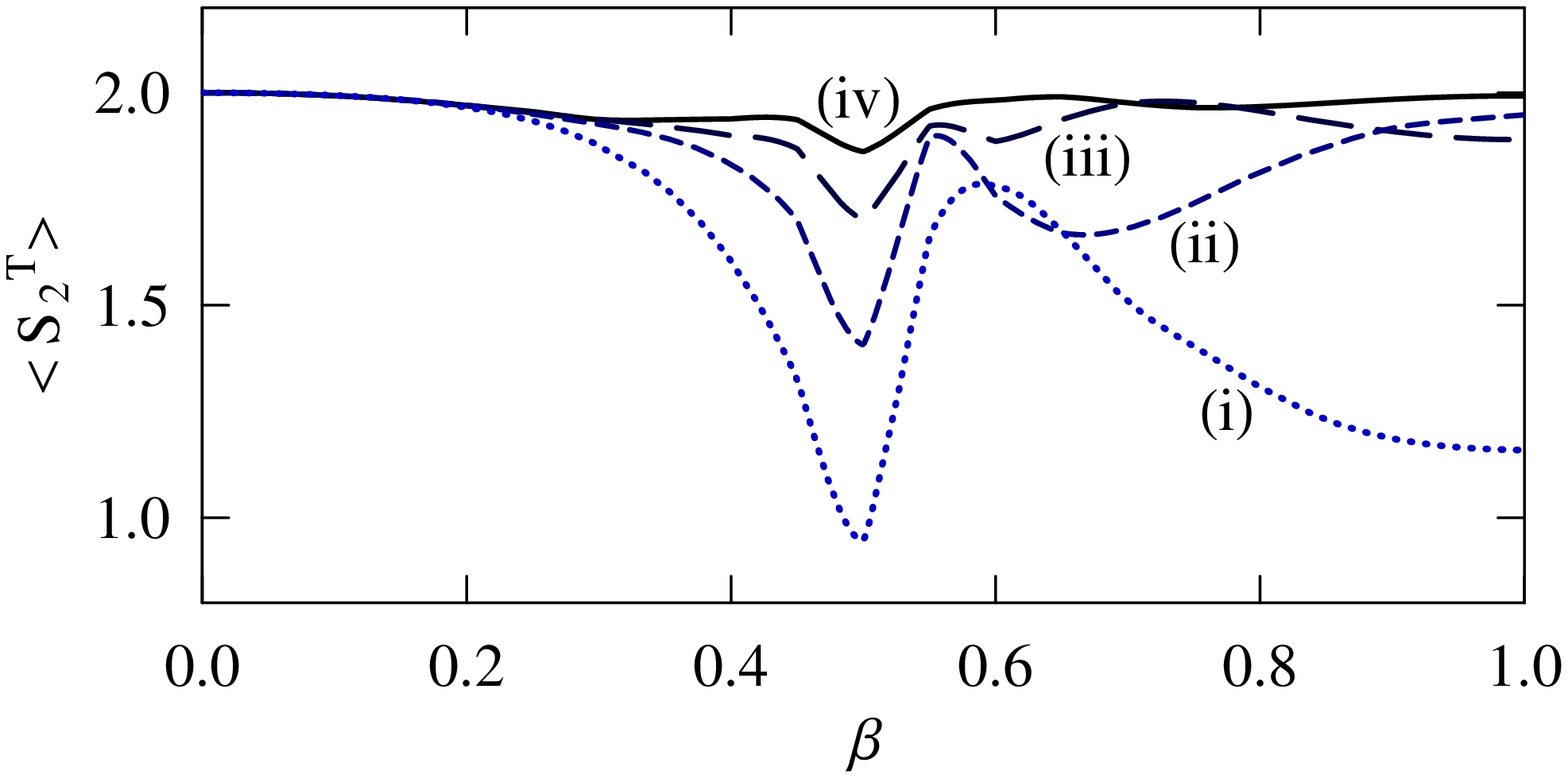}
\caption{(Color online) Exact long-term time average of $S_2^T$ as a
function of $\beta = \lambda / (1 + \lambda)$ for various system
sizes: $N = 40$ and $D = 3d = 6$ (i); $N = 60$ and $D = 3d = 9$
(ii); $N = 80$ and $D = 3d = 12$ (iii); $N = 100$ and $D = 3d = 15$
(iv). \label{fig-3}}
\end{figure}

The long-term time average of the topological R\'{e}nyi entropy
against the magnetic field $\lambda$ of the quench Hamiltonian
$\hat{H} (\lambda)$ is plotted in Fig. \ref{fig-3}. The time
averages are calculated from $1000$ independent time instants, which
are taken from a sufficiently long interval at sufficiently large
times such that $t \gg \max \{ 1, \lambda^{-1} \}$ for all time
instants $t$. We first notice that $\langle S_2^T \rangle$ is
distinct from the value of $S_2^T = 0$ that would correspond to the
lack of topological order. Instead, it is found for all magnetic
fields $\lambda$ that $\langle S_2^T \rangle$ is close to the
initial value of $S_2^T = 2$ and gets closer to it if we increase
the system size $N$ as well as the subsystem dimensions $D$ and $d$.
We therefore argue that $S_2^T = 2$ in the thermodynamic limit for
all times $t$ after a quench with an arbitrary magnetic field
$\lambda$.

\emph{Perturbation theory.---}In the case of $\kappa = 1$, the
Hamiltonian in Eq. (\ref{eq-H-2}) requires perturbation theory
around the exactly solvable point at $\lambda = 0$. Our aim is to
determine the time evolution of the first-order correction to
$S_2^T$. In general, the most straightforward approach is the method
of perturbative continuous unitary transformations \cite{Halasz,
Dusuel, PCUT}. However, a naive first-order perturbation theory is
sufficient in our case, and we derive our results using this simpler
approach.

The unperturbed ground state of the initial Hamiltonian $\hat{H}
(0)$ is $| 0 \rangle$ and the perturbed ground state of the quench
Hamiltonian $\hat{H} (\lambda)$ is $| \Omega \rangle$. In the first
order of perturbation theory, the perturbed ground state takes the
expansion form $| \Omega \rangle = | 0 \rangle + (\lambda / 4)
\sum_{\langle \mathbf{r_1}, \mathbf{r_2} \rangle} | \{ \mathbf{r_1},
\mathbf{r_2} \} \rangle$, where $| \{ \mathbf{r_1}, \mathbf{r_2} \}
\rangle$ is the state with two star excitations at the positions
$\mathbf{r_1}$ and $\mathbf{r_2}$ such that $A_{\mathbf{r_1}}^z =
A_{\mathbf{r_2}}^z = -1$. The position of each star is labeled by
the 2D vector $\mathbf{r} = (x, y)$ with $0 \leq x,y < N$. In terms
of the ground state $| \Omega \rangle$, the initial state $| \Psi(0)
\rangle$ becomes $| \Psi(0) \rangle = | 0 \rangle = | \Omega \rangle
- (\lambda / 4) \sum_{\langle \mathbf{r_1}, \mathbf{r_2} \rangle} |
\{ \mathbf{r_1}, \mathbf{r_2} \} \rangle$. Although the
non-degenerate corrections to the excited states $| \{ \mathbf{r_1},
\mathbf{r_2} \} \rangle$ vanish in the first-order calculation, the
perturbation introduces a hopping between these degenerate states.
The two star excitations can hop between neighboring stars with an
amplitude $-\lambda$, and their only interaction is a hard-core
repulsion: they are not allowed to be at the same star. However,
this interaction is negligible in the thermodynamic limit because
the two excitations are far away from each other. The exact
eigenstates of the quench Hamiltonian with two star excitations are
then $|\mathbf{k_1}, \mathbf{k_2} \rangle = N^{-2}
\sum_{\mathbf{r_1}, \mathbf{r_2}} e^{i (\mathbf{K_1} \cdot
\mathbf{r_1} + \mathbf{K_2} \cdot \mathbf{r_2})} | \{ \mathbf{r_1},
\mathbf{r_2} \} \rangle$, where $\mathbf{k} = (k_x, k_y)$ with $0
\leq k_x, k_y < N$, and each excitation has a 2D momentum
$\mathbf{K} = 2\pi \mathbf{k} / N$. Note that the state $| \{
\mathbf{r_1}, \mathbf{r_2} \} \rangle = | \{ \mathbf{r_2},
\mathbf{r_1} \} \rangle$ appears twice in the sum. Since the system
and the initial state are both invariant under translations, we only
need to consider the eigenstates with zero total momentum
$\mathbf{k_1} + \mathbf{k_2}$. These states labeled by $| \mathbf{k}
\rangle \equiv | \mathbf{k}, -\mathbf{k} \rangle$ have relative
energies $E(\mathbf{k}) = 4 - 4\lambda \epsilon(\mathbf{k})$ with
respect to the ground state $| \Omega \rangle$, where
$\epsilon(\mathbf{k}) \equiv \cos (K_x) + \cos (K_y)$. In terms of
the eigenstates, the initial state $| \Psi(0) \rangle$ takes the
form $| \Psi(0) \rangle = | 0 \rangle = | \Omega \rangle - (\lambda
/ 2) \sum_{\mathbf{k}}' \epsilon(\mathbf{k}) | \mathbf{k} \rangle$,
where the prime means that the summation is only over half of the
$N^2$ possible $\mathbf{k}$ values. The state $| \Psi(t) \rangle$
after the time evolution with $\hat{H} (\lambda)$ is then obtained
by the substitution $\epsilon(\mathbf{k}) \rightarrow
\epsilon(\mathbf{k}) e^{-i t E(\mathbf{k})}$, and in terms of the
states $| \{ \mathbf{r_1}, \mathbf{r_2} \} \rangle$ it reads
\begin{equation}
| \Psi(t) \rangle = | \Omega \rangle + \sum_{(\mathbf{r_1},
\mathbf{r_2})} Q(\mathbf{r_1}, \mathbf{r_2}, t) | \{ \mathbf{r_1},
\mathbf{r_2} \} \rangle, \label{eq-psi}
\end{equation}
where $(\mathbf{r_1}, \mathbf{r_2})$ means that the summation is
over pairs of stars without double counting. The coefficient of each
state is $Q(\mathbf{r_1}, \mathbf{r_2}, t) = - (\lambda / N^2)
\sum_{\mathbf{k}}' \epsilon(\mathbf{k}) \cos [\mathbf{K} \cdot
(\mathbf{r_1} - \mathbf{r_2})] e^{-i t E(\mathbf{k})}$. Since the
functions $Q(\mathbf{r_1}, \mathbf{r_2}, t)$ consist of many
incoherent oscillations with different frequencies $E(\mathbf{k})$,
the average modulus square of any such function is given by
\begin{equation}
\big{\langle} | Q(\mathbf{r_1}, \mathbf{r_2}, t) |^2 \big{\rangle} =
\frac{\lambda^2}{N^4} \sum_{\mathbf{k}}' \epsilon(\mathbf{k})^2
\cos^2 \big{[} \mathbf{K} \cdot (\mathbf{r_1} - \mathbf{r_2})
\big{]} = \frac{\lambda^2}{4N^2} \, . \label{eq-Q}
\end{equation}
This result has a simple physical interpretation. The two
excitations are always at neighboring stars before the time
evolution, and the sum of the norm squares in the states with two
excitations is $2N^2 \lambda^2 / 16 = N^2 \lambda^2 / 8$ because
there are $2N^2$ ways of choosing two neighboring stars. During the
time evolution, the excitations hop between stars, and the norm
square is distributed uniformly between all possible states with two
star excitations. Since there are $N^2 (N^2-1) / 2 \approx N^4 / 2$
ways of choosing any two stars, the uniform norm square corresponds
to the result $\lambda^2 / 4N^2$ in Eq. (\ref{eq-Q}).

When calculating the R\'{e}nyi entropy for the state $| \Psi(t)
\rangle$ in Eq. (\ref{eq-psi}), there are two corrections to the
value $S_2^{\partial A} = L - n$ for the unperturbed ground state $|
0 \rangle$: a static correction from the perturbed ground state $|
\Omega \rangle$ and a dynamic correction from the oscillating terms
$\propto Q(\mathbf{r_1}, \mathbf{r_2}, t)$. The static correction is
linearly proportional to the combined length $L$ of the boundary in
each case $(m)$, therefore its topological contribution vanishes
\cite{Halasz}. On the other hand, the terms $\propto Q(\mathbf{r_1},
\mathbf{r_2}, t)$ give an expectation value $\langle \Psi(t) |
\hat{A}_{\mathbf{r_1}}^x \hat{A}_{\mathbf{r_2}}^x | \Psi(t) \rangle
= 2 \, \mathrm{Re} [Q(\mathbf{r_1}, \mathbf{r_2}, t)]$ for each pair
of stars $\mathbf{r_1}$ and $\mathbf{r_2}$ on the same closed loop
of the boundary. Since $\langle \textrm{Re} [Q(\mathbf{r_1},
\mathbf{r_2}, t)]^2 \rangle = \langle |Q(\mathbf{r_1}, \mathbf{r_2},
t)|^2 \rangle/2$ and there are $\ell (\ell - 1)/2$ ways of choosing
two stars from a closed loop of length $\ell$, the time average of
the dynamic correction is $\langle \Delta S_2^{\partial A} \rangle =
-\lambda^2 \ell (\ell - 1) / (4N^2 \ln 2)$ for each closed loop of
the boundary in each case $(m)$. Since there is a term $\propto
\ell^2$, the topological contribution does not vanish, and the
resulting time average of the topological R\'{e}nyi entropy is
\begin{equation}
\langle S_2^T \rangle = 2 - \frac{\lambda^2 \left( 8D^2 - 16d^2
\right)} {N^2 \ln 2} \, . \label{eq-S-1}
\end{equation}
If we take both the system size and the subsystem dimensions to
infinity such that $D \sim d \sim N^{1/4}$, the perturbative
correction vanishes as $N^{-3/2}$. We can thus choose macroscopic
subsystems in the thermodynamic limit such that the topological
R\'{e}nyi entropy is $S_2^T = 2$ for all times $t$.

\begin{figure}[t!]
\centering
\includegraphics[width=8.0cm]{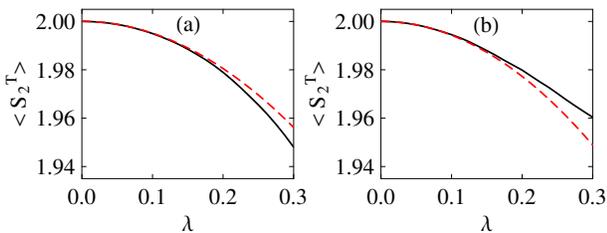}
\caption{(Color online) Long-term time average of $S_2^T$ from the
exact treatment (solid black line) and the first-order perturbation
theory (dashed red line) for two system sizes: $N = 40$ and $D = 3d
= 6$ (a); $N = 80$ and $D = 3d = 12$ (b). \label{fig-4}}
\end{figure}

To verify that the first-order perturbation theory is indeed a good
approximation at $\lambda \ll 1$ and is not invalidated by
higher-order corrections, we also establish an analogous
perturbation theory in the exactly solvable $\kappa = 0$ case. The
first-order time average of the topological R\'{e}nyi entropy takes
the form
\begin{equation}
\langle S_2^T \rangle = 2 - \frac{\lambda^2 \left( d^2 + 6D - 6d - 1
\right)} {2N \ln 2} \, . \label{eq-S-2}
\end{equation}
Note that the correction in this case vanishes as $N^{-1/2}$ when $D
\sim d \sim N^{1/4}$. The exact results and those obtained from the
perturbation theory are plotted together in Fig. \ref{fig-4}. We
notice that the respective curves are in good agreement when
$\lambda \ll 1$, and therefore we argue that the first-order
perturbation theory is a good approximation in the $\kappa = 1$ case
as well.

\emph{Conclusions.---}In this Letter, we studied the resilience of
topological order in the TCM ground state after a quantum quench
with an external magnetic field. The time evolution of topological
order was detected via the topological R\'{e}nyi entropy of order
$2$. We considered two different quenches: the integrable one was
solved exactly, while the non-integrable one was treated with
perturbation theory. In both cases, we found that topological order
is resilient in the thermodynamic limit. Our results are in
agreement with those in Ref. \cite{Tsomokos}, but they apply to
significantly larger system sizes.

It is interesting to discuss the generic conditions under which
topological order is resilient. First of all, the results for the
non-integrable quench show that integrability does not play a role.
On the other hand, both quenches preserve the $\mathbb{Z}_2$ gauge
structure of the TCM ground state. This gauge structure allows us to
use the modified definition for $S_2^T$, which in turn makes the
subsequent calculations possible. We therefore believe that
topological order characterized by $S_2^T$ is a robust
non-equilibrium property of the system whenever gauge invariance is
preserved. In perspective, it would be important to study the
behavior of topological order after a more generic
non-gauge-preserving quench. This further step could help us better
understand the universal non-equilibrium and thermal properties of
topologically ordered systems.

We thank X.-G. Wen for illuminating discussions. This work was
supported in part by the National Basic Research Program of China
Grants No. 2011CBA00300 and No. 2011CBA00301 and the National
Natural Science Foundation of China Grants No. 61073174, No.
61033001, and No. 61061130540. Research at the Perimeter Institute
for Theoretical Physics is supported in part by the Government of
Canada through NSERC and by the Province of Ontario through MRI.

%%%%%%%%%%%%%%%%%%%%%%%%%%%%%%%%%%%%%%%%%%%%%%%%

%%%%%%%%%%%%%%%%%%%%%%%%%%%%%%%%%%%%%%%%%%%%%%%%

\end{document}